\documentstyle[twoside,12pt]{article}

\catcode`\@=11
\long\def\@makefntext#1{ 
\protect\noindent \hbox to 3.2pt {\hskip-.9pt
$^{{\eightrm\@thefnmark}}$\hfil}#1\hfill} 

\def\thefootnote{\fnsymbol{footnote}}
 \def\@makefnmark{\hbox to 0pt{$^{\@thefnmark}$\hss}}  

\def\ps@myheadings{\let\@mkboth\@gobbletwo
\def\@oddhead{\hbox{} 
\rightmark\hfil\eightrm\thepage}
\def\@oddfoot{}\def\@evenhead{\eightrm\thepage\hfil 
\leftmark\hbox{}}\def\@evenfoot{}
\def\sectionmark##1{}\def\subsectionmark##1{}}

\oddsidemargin=\evensidemargin
\renewcommand{\thefootnote}{\fnsymbol{footnote}}

\newcounter{sectionc}\newcounter{subsectionc}\newcounter{subsubsectionc}
\renewcommand{\section}[1] {\vspace{12pt}\addtocounter{sectionc}{1}
\setcounter{subsectionc}{0}\setcounter{subsubsectionc}{0}\noindent
	{\bf\thesectionc. #1}\par\vspace{5pt}}
\renewcommand{\subsection}[1] {\vspace{12pt}\addtocounter{subsectionc}{1}
	\setcounter{subsubsectionc}{0}\noindent
	{\bf\thesectionc.\thesubsectionc. {\kern1pt \bf\it #1}}\par\vspace{5pt}}
\renewcommand{\subsubsection}[1] {\vspace{12pt}\addtocounter{subsubsectionc}{1}
	\noindent{\thesectionc.\thesubsectionc.\thesubsubsectionc.
	{\kern1pt \it #1}}\par\vspace{5pt}}
\newcommand{\nonumsection}[1] {\vspace{12pt}\noindent{\bf #1}
	\par\vspace{5pt}}

\newcommand{\textlineskip}{\baselineskip=14pt}
\newcommand{\smalllineskip}{\baselineskip=12pt}

\def\eightcirc{
\begin{picture}(0,0)
\put(4.4,1.8){\circle{6.5}}
\end{picture}}
\def\eightcopyright{\eightcirc\kern2.7pt\hbox{\eightrm c}}

\def\abstracts#1#2#3{{
	\centering{\begin{minipage}{5in}\baselineskip=12pt\tenrm
	\centerline{ABSTRACT}
	\parindent=0pt #1\par
	\parindent=15pt #2\par
	\parindent=15pt #3
	\end{minipage} }\par}}

\renewenvironment{thebibliography}[1]			
	{
	 \begin{list}{\arabic{enumi}.}			
	{\usecounter{enumi}\setlength{\parsep}{0pt}
	 \setlength{\leftmargin 17pt}{\rightmargin 0pt}	
	 \setlength{\itemsep}{0pt} \settowidth		
	{\labelwidth}{#1.}\sloppy}}{\end{list}}	

\newcounter{itemlistc}
\newcounter{romanlistc}
\newcounter{alphlistc}
\newcounter{arabiclistc}

\newcommand{\tcaption}[1]{                      
        \addtocounter{table}{1}
         {{\tenrm\offinterlineskip Table~\thetable . #1} }\hfil\break }

\def\pmb#1{\setbox0=\hbox{#1}
	\kern-.025em\copy0\kern-\wd0
	\kern.05em\copy0\kern-\wd0
	\kern-.025em\raise.0433em\box0}

\def\fnt#1#2{\footnotetext{\kern-.3em
	{$^{\mbox{\scriptsize #1}}$}{#2}}}

\def\fpage#1{\begingroup
\voffset=.3in
\thispagestyle{empty}\begin{table}[b]\centerline{\footnotesize #1}
	\end{table}\endgroup}

 1
 1
 1

\font\eightrm=cmr8

\def\qed{\hbox{${\vcenter{\vbox{                          
   \hrule height 0.4pt\hbox{\vrule width 0.4pt height 6pt
   \kern5pt\vrule width 0.4pt}\hrule height 0.4pt}}}$}}

\begin{document}
\normalsize\textlineskip
{\thispagestyle{empty}
\setcounter{page}{1}

\renewcommand{\thefootnote}{\fnsymbol{footnote}} 
\def\bsc{{\sc a\kern-6.4pt\sc a\kern-6.4pt\sc a}}
\def\bflatex{\bf L\kern-.30em\raise.3ex\hbox{\bsc}\kern-.14em
T\kern-.1667em\lower.7ex\hbox{E}\kern-.125em X}

\fpage{1}
\rightline{UNITU-THEP-12/1993}
\rightline{hep-ph/9310293}
\centerline{\bf THE PROTON SPIN STRUCTURE IN SKYRME TYPE MODELS
$^\dagger$}
\vspace{0.37truein}
\centerline{\footnotesize HERBERT WEIGEL}
\vspace*{0.015truein}
\centerline{\footnotesize\it Institut for Theoretical Physics,
T\"ubingen University}
\baselineskip=12pt
\centerline{\footnotesize\it D-72076 T\"ubingen, FR Germany}

\vspace*{0.21truein}

\abstracts{It is demonstrated how in the soliton approach to baryons
a small but non-vanishing matrix element of the axial singlet current
is obtained. Furthermore it is shown that these models also provide a
mechanism to disentangle the ``matter" and ``glue" contributions to
this matrix element. Numerical results indeed exhibit a cancellation
between these two components. However, the magnitudes of both turn out
to be significantly smaller than unity.}{}{}

\vspace*{0.21truein}\textlineskip

\noindent
It has been suggested that the small matrix element\cite{emc} of the
axial singlet current $J_\mu^5$ between nucleon states is due to a
cancellation between ``matter" and ``glue" contributions\cite{efr88}.
Such an explanation is motivated by the anomaly equation:
\begin{eqnarray}
\partial^\mu J_\mu^5=2i\sum_i m_i {\bar q}_i\gamma_5 q_i
+\partial^\mu K_\mu
\label{anomaly}
\end{eqnarray}

\noindent
with $K_\mu$ being the Chern-Simons current of QCD. Due to gauge variance
the ``glue" contribution cannot straightforwardly be linked to the matrix
element of $K_\mu$. A gauge invariant decomposition can, however, be
formulated in terms of three-point functions for the coupling of the
pseudoscalar field $\eta^\prime$ and the glueball field $G=\partial^\mu
K_\mu$ to nucleons\cite{sh90}. The main goal of this talk is to
summarize the work of refs.\cite{sch90,sch92,sch93} where it has
been demonstrated that a decomposition
can also be formulated in models which do not contain baryons explicitly
but rather have baryons emerging as solitons of an effective meson theory.
In ref.\cite{sch93} an eleborate meson action has been developed to describe
two- and three-point functions for mesons at tree level. Let $U$ denote
the non-linear realization of the pseudoscalar mesons. Then the axial anomaly
(\ref{anomaly}) is incorporated by coupling $G$ to the pseudoscalar singlet
field $\chi=(-i/3){\rm log\ det}\ U$. Elimination of $G$ via its equation
of motion yields a massive $\eta^\prime$. Furthermore the Lagrangian which
already contains the relevant $SU(3)$ symmetry breaking terms is augmented
by OZI-violating terms in the pseudoscalar singlet channel. Then not only the
empirical values for the masses of the $\eta$ and $\eta^\prime$ mesons are
reproduced but also the widths of their decays into two photons. The soliton
solutions of this effective meson theory acquire the hedgehog shape:
\begin{eqnarray}
U_0({\bf r})={\rm exp}\big(i{\bf \hat r}\cdot
{\mbox{\boldmath $\tau$}} F(r)\big),\quad
\omega_0={{\omega(r)}\over{2g}},\quad
\rho_{i,a}={{G(r)}\over{gr}}\epsilon_{ija}\hat r_j.
\label{hedgehog}
\end{eqnarray}

\noindent
while all other fields, especially the $\eta$-fields, vanish
in the static limit.

\vfill
\noindent
------------------------- \hfil\break
{\footnotesize
$^\dagger $ Invited talk presented at the 13$^{\rm th}$ PANIC
meeting, Perugia, June 28$^{\rm th}$-July 3$^{\rm nd}$ 1993.}
\eject

The radial functions $F(r),\omega(r)$ and $G(r)$
are obtained by minimizing the static energy functional. This field
configuration is quantized by introducing collective coordinates for
the rotations in $SU(3)$ flavor space
\begin{eqnarray}
U({\bf r},t)=A(t)U_0({\bf r})A^\dagger(t), \qquad A(t)\in SU(3)
\label{coll}
\end{eqnarray}

\noindent
and similarly for the vector fields. A special feature of the collective
rotation is the excitation of field components which vanish in the static
case. As an example the pseudoscalar isoscalar fields are displayed:
\begin{eqnarray}
\eta_0\ {\bf 1}_{3\times3}+\eta_8\lambda_8=
\pmatrix{\hspace{8pt} \eta_T(r){\bf \hat r}
\cdot{\bf \Omega}\ {\bf 1}_{2\times2}&
{| \atop |}&\hspace{-10pt}{0\atop 0}\cr
-------\ -\hspace{-8pt}&-&\hspace{-10pt}-----\cr
0\qquad 0&|&\hspace{-8pt}\eta_S(r){\bf \hat r}\cdot{\bf \Omega}\cr}.
\label{eta}
\end{eqnarray}

\noindent
The anglar velocities describe the time dependence of the
collective rotations:
$A^\dagger{\dot A}=(i/2){\sum^8_{a=1}}{\lambda_a}{\Omega_a}.$
Extremization of the moments of inertia determines the radial functions
$\eta,\eta^\prime$, etc.. The collective Hamiltonian which also contains
$SU(3)$ symmetry breaking terms is diagonalized exactly. The resulting
eigenstates are identified as baryons. Numerically reasonable mass
differences between baryons with different spin and/or hypercharge are
obtained, {\it i.e.} the model reproduces the empirical data for the mass
differences with in 15\%\cite{pa91}. The non-electromagnetic part
of the neutron-proton mass difference is evaluated using a perturbation
expansion in the isospin breaking $y=(m_u-m_d)/(m_u+m_d)$ and obtained to
be $(M_n-M_p)_{\rm non-em}=y\Delta$. The integral $\Delta$ involves
the meson fields:
\begin{eqnarray}
\Delta=\frac{\pi}{3} m_\pi^2 F_\pi^2 \int dr r^2
{\rm sin}F(r) \eta_T(r)+\cdot\cdot\cdot
\label{delta}
\end{eqnarray}

\noindent
where only one characteristic term is displayed. The analysis of the meson
sector yields $y\approx-0.42$. The axial singlet current $J_\mu^5$ is
obtained to be
\begin{eqnarray}
J_\mu^5=-2\frac{\partial{\cal L}}{\partial(\partial^\mu\chi)}
=\sqrt3 F_\pi \partial_\mu \eta^\prime + s \tilde J_\mu^5.
\label{jm5}
\end{eqnarray}

\noindent
The ``fudge factor" $s$ has been introduced to allow for deviations from
the nonet structure\cite{sch90}. $\tilde J_\mu^5$ stems from the anomalous
parts of the action and is only non-vanishing when vector meson fields are
present. The matrix element of $J_\mu^5$ between proton states defines
two momentum dependent form factors
\begin{eqnarray}
\langle P({\bf p}\ ^\prime)|J_\mu^5|P({\bf p})\rangle=
{\overline u}({\bf p}\ ^\prime)\big[\gamma_\mu\gamma_5H(q^2)+
{{i q_\mu}\over{2M_p}}\gamma_5\tilde H(q^2)\big]u({\bf p})
\qquad {\rm with}\ q_\mu=p_\mu^\prime-p_\mu
\label{formf}
\end{eqnarray}

\noindent
The main form factor $H(0)$ is identified as the quarks' contribution
to the proton spin. Employing the Dirac equation and noting that the
induced form factor $\tilde H(q^2)$ has no pole at $q=0$ since
$\eta^\prime$ is massive one finds for $H(0)$:
\begin{eqnarray}
\langle P({\bf p})|s \partial^\mu \tilde J_\mu^5
|P({\bf p})\rangle=
2M_p H(0) {\overline u}({\bf p})\gamma_5u({\bf p}).
\label{h0}
\end{eqnarray}

\noindent
One recognizes that the quarks' contribution to the
proton spin vanishes when $\tilde J_\mu^5=0$ {\it i.e.} no short range
effects (vector mesons or explicit quarks) are present.
The coupling of the glueball field to the nucleon is introduced
in a chirally invariant manner
\begin{eqnarray}
{\cal L}_{GNN}=\frac{2t}{m_{\eta^\prime}^4}\partial^\mu G\tilde J_\mu^5
\end{eqnarray}

\noindent
The equations of motion for $\eta^\prime$ and $G$ allow to extract
their couplings to the nucleon:
\begin{eqnarray}
g_{\eta^\prime NN}=\frac{s-t}{s}\frac{2M_p}{\sqrt3 F_\pi} H(0)
\qquad {\rm and}\qquad g_{GNN}=\frac{t}{t-s}
\frac{2\sqrt3 F_\pi}{m_{\eta^\prime}^4}g_{\eta^\prime NN}.
\label{cconst}
\end{eqnarray}

\noindent
Eliminating the parameters $s$ and $t$ finally gives the
desired decomposition:
\begin{eqnarray}
H(0)=\frac{\sqrt3 F_\pi}{2M_p}\left(g_{\eta^\prime NN}
-\frac{m_{\eta^\prime}^4}{2\sqrt3 F_\pi}g_{GNN}\right)
={\rm ``matter"}+{\rm ``glue"}.
\label{decomp}
\end{eqnarray}

\noindent
\begin{table}
\tcaption{The decomposition of the axial singlet matrix element
$H(0)$ into ``matter" and ``glue" contributions. `baryon best fit' and
`meson best fit' refer to parameter sets which are fixed in the baryon
and meson sectors\cite{sch93}. The errors are due to the uncertainty in
$(M_n-M_p)_{\rm non-em}$.}
\newline
\centerline{\tenrm\smalllineskip
\begin{tabular}{l||c|c||c|c}
&\multicolumn{2}{c||}{\rm baryon best fit}
&\multicolumn{2}{c}{\rm meson best fit} \\
\hline
\hline
$H(0)$ & {\rm ``matter"} & {\rm ``glue"}
& {\rm ``matter"} & {\rm ``glue"} \\
\hline
{}~~0.0&$0.54\pm0.26$&$-0.54\pm0.26$&$0.32\pm0.26$&$-0.32\pm0.26$ \\
\hline
{}~~0.2&$0.58\pm0.30$&$-0.38\pm0.30$&$0.31\pm0.26$&$-0.11\pm0.28$ \\
\hline
{}~~0.4&$0.58\pm0.28$&$-0.18\pm0.28$&$0.32\pm0.27$&$~~~0.08\pm0.27$ \\
\end{tabular}}
\end{table}
It is now possible to evaluate the ``matter" and ``glue" contribution
to the axial singlet current as a function of its magnitude. {\it I.e.}
we consider $H(0)$ as a parameter fixing $s$. Furthermore $t$ is
adjusted to reproduce $(M_n-M_p)_{\rm non-em}$ which is most
sensitive to the $\eta$ meson profiles. The numerical results are
displayed in table 1. These results reveal two major facts. Firstly,
we observe that there is indeed a cancellation between the ``matter"
and ``glue" contributions to $H(0)$. However, secondly, the individual
magnitudes  are significantly smaller than unity. Thus the interpretation
that the $U_A(1)$ anomaly (\ref{anomaly}) is responsible for the small
axial singlet matrix is not supported by Skyrme type models.

\nonumsection{Acknowledgements}

The author would like to thank his colleagues J. Schechter and
A. Subbaraman  who participated in the work underlying this talk.
This work is supported by the Deutsche Forschungsgemeinschaft (DFG) under
contract Re 856/2-1.

\vfill\eject

\nonumsection{References}

\end{document}